\documentclass[11pt]{article}
\usepackage{amsmath,epsf,amssymb,latexsym,enumerate,cite,shadow,array,color}
\usepackage[ps,dvips,matrix,arrow,frame,import,curve,color]{xy}
\renewcommand{\theequation}{\thesection.\arabic{equation}}
\newcounter{saveeqn}
\newcommand{\add}{\addtocounter{equation}{1}}
\newcommand{\alpheqn}{\setcounter{saveeqn}{\value{equation}}%
\setcounter{equation}{0}%
\renewcommand{\theequation}{\mbox{\thesection.\arabic{saveeqn}{\alph{equation}}}}}
\newcommand{\reseteqn}{\setcounter{equation}{\value{saveeqn}}%
\renewcommand{\theequation}{\thesection.\arabic{equation}}}

\newif\iffigs\figstrue

\DeclareFontFamily{U}{rsf}{}
\DeclareFontShape{U}{rsf}{m}{n}{
  <5> <6> rsfs5 <7> <8> <9> rsfs7 <10-> rsfs10}{}
\DeclareMathAlphabet\Scr{U}{rsf}{m}{n}

\def\pplogo{\vbox{\kern-\headheight\kern -29pt
\halign{##&##\hfil\cr&{
\ppnumber}\cr\rule{0pt}{2.5ex}&\ppdate\cr}
}}
\makeatletter
\def\ps@firstpage{\ps@empty \def\@oddhead{\hss\pplogo}%
  \let\@evenhead\@oddhead 
}
\def\maketitle{\par
 \begingroup
 \def\thefootnote{\fnsymbol{footnote}}
 \def\@makefnmark{\hbox{$^{\@thefnmark}$\hss}}
 \if@twocolumn
 \twocolumn[\@maketitle]
 \else \newpage
 \global\@topnum\z@ \@maketitle \fi\thispagestyle{firstpage}\@thanks
 \endgroup
 \setcounter{footnote}{0}
 \let\maketitle\relax
 \let\@maketitle\relax
 \gdef\@thanks{}\gdef\@author{}\gdef\@title{}\let\thanks\relax}
\makeatother

\newcommand{\bea}{\begin{eqnarray}}
\newcommand{\eea}{\end{eqnarray}}
\newcommand{\be}{\begin{equation}}
\newcommand{\ee}{\end{equation}}

\def\O{\Scr{O}}

\def\cQ{{\Scr Q}}

\def\cV{{\Scr V}}
\def\cU{{\Scr U}}

\def\cA{{\Scr A}}
\def\cB{{\Scr B}}

\def\cC{{\Scr C}}
\def\cD{{\Scr D}}

\def\cP{{\Scr P}}

\begin{document}

\thispagestyle{empty}

\begin{titlepage}
\begin{flushright}
SU-ITP-2008-04
\end{flushright}


\vskip  2 cm

\vspace{24pt}

\begin{center}
{ \LARGE \textbf{ Explicit Action of  $E_{7(7)}$  on  ${\cal{N}}=8$ Supergravity Fields }}

\vspace{28pt}

{\bf Renata Kallosh}\footnote{\normalsize{\ kallosh@stanford.edu}} \hspace{0.2cm} and \hspace{0.2cm}{\bf
Masoud Soroush}\footnote{\normalsize{\ soroush@stanford.edu}}

    \vspace{15pt}

 \textsl{Department of Physics,
    Stanford University}\\ \textsl{Stanford, CA 94305-4060, USA}

\vspace{10pt}

\vspace{24pt}

\end{center}

\begin{abstract}

We present an explicit,   exact to all orders in gravitational coupling  $E_{7(7)}$ symmetry transformations of on-shell  ${\cal{N}}=8$ supergravity fields in the gauge with 70 scalars in ${E_{7(7)}\over SU(8)}$ coset space,  the local $SU(8)$ symmetry being fixed. The non-linear realization of $E_{7(7)}$ includes a field-dependent $SU(8)$ transformation preserving the unitary gauge. We find the conserved Noether-Gaillard-Zumino  current of $E_{7(7)}$ symmetry, the linear part of it being a chiral $SU(8)$ symmetry. We comment on the conformal realization of the  $E_{7(7)}$  algebra  which includes a dilatation operator.  We hope that these results can be useful for studies of anomalies/absence of anomalies and the  UV behavior of ${\cal{N}}=8$ supergravity.

\end{abstract}

\end{titlepage}
\newpage

\section{Introduction}

The revival of the interest to ${\cal{N}}=8$  supergravity is due to the recent computations of the loop corrections, which reveal the 3-loop finiteness and indicate a possibility that ${\cal{N}}=8$ supergravity may be UV finite \cite{Bern:2007hh,Bern:2006kd} or at least seem to have some very interesting and unexpected features
\cite{Mason:2007ct} - \cite{BjerrumBohr:2008vc}. The no-triangle hypothesis  of the one-loop graphs conjectured in the framework of helicity amplitudes in \cite{BjerrumBohr:2006yw}  was tested in many cases and may be related to  the all-loop finiteness. It was suggested in \cite{Kallosh:2007ym} that using the manifestly ${\cal{N}}=8$  supersymmetric background field method may add some additional information on the properties of quantum corrections of ${\cal{N}}=8$ supergravity and that the background field theory version of the no-triangle property of this theory may be related to the absence of anomalies. For example, one can think about  the classical local $SU(8)$  symmetry: it  has 63 generators, 28 generators of the $SO(8)$ and 35 orthogonal to $SO(8)$. The 35 symmetries act with the opposite sign on the left and right chiral projections of the fermion fields and potentially may lead to $\gamma^5$-anomaly. It is not clear if this symmetry is anomalous or not and whether the presence/absence of anomalies of the local $SU(8)$   is relevant to the UV properties of ${\cal{N}}=8$ supergravity. The Weyl anomalies in extended supergravities have been studied before and are reviewed in \cite{Duff:1993wm}.

To understand better  the  anomalies in ${\cal{N}}=8$ supergravity we would like to specify the classical symmetries which may or may not be broken by quantum corrections. This is the purpose of this paper with regard to  the  ``hidden'' $E_{7(7)}$ symmetry .

Classical ${\cal{N}}=8$ supergravity has in addition to 8 local supersymmetries also a local $SU(8)$ symmetry and a hidden global $E_{7(7)}$ symmetry \cite{Cremmer:1978ds} -\cite{de Wit:1982ig} on shell, when the exact non-linear equations of motion are satisfied. The early versions of ${\cal{N}}=8$  supergravity had a rigid $SO(8)$ symmetry \cite{deWit:1977fk}  and a rigid $SU(8)$  symmetry \cite{deWit:1978sh}. It was recognized in \cite{Cremmer:1978ds}, \cite{Cremmer:1979up} that when the maximal extended  four-dimensional supergravity was derived from the eleven-dimensional one, the theory had a  gauge $SU(8)$ symmetry and a hidden rigid  $E_{7(7)}$ symmetry.

The $E_{7(7)}$ symmetry is realized linearly and independently from the local $SU(8)$ symmetry and it acts on 133 scalars present in the classical action before gauge-fixing as well as on the vectors of the theory. The gauge-fixing can use the 63 local parameters of $SU(8)$  to remove 63 non-physical scalars so that only 70 physical  scalars are left. This leads to a non-linear realization of the $E_{7(7)}$ on the remaining 70 fields.  The  $E_{7(7)}$ transformation has to be performed simultaneously with the gauge preserving field dependent $SU(8)$ transformation.  ${\cal{N}}=8$ supergravity  without and with $SU(8)$ gauge-fixing  was derived and studied in  \cite{Cremmer:1978ds,Cremmer:1979up,Gaillard:1981rj,de Wit:1982ig}.

Here we will present a useful form of the non-linear realization of $E_{7(7)}$ symmetry including the accompanied  transformation preserving the unitary gauge. The interest to such study is motivated by the recent proposals that ${\cal{N}}=8$ supergravity may be UV finite \cite{Bern:2007hh,Bern:2006kd}.  It is also interesting that the computations have been performed mostly using the ${\cal{N}}=4$ SYM theory, using  the  KLT relation \cite{Kawai:1985xq}
 between ${\cal{N}}=4$ SYM theory and ${\cal{N}}=8$ supergravity which states that at the tree level the product of two copies of gauge theory can define supergravity S-matrix. A symmetry of each ${\cal{N}}=4$ SYM theory is $SU(4)$ and two of them will give $SU(4)\times SU(4)$ but where is the rest coming? This puzzle is currently under investigation \cite{BEF}. With regard to the diagonal subgroup of  $E_{7(7)}$ symmetry the problem can be reduced to the rigid $SU(8)$ symmetry of the spectrum of the physical states of ${\cal{N}}=8$ supergravity with 63 parameters. However, the orthogonal 70 symmetries of $E_{7(7)}$ are realized non-linearly and the origin of these symmetries from the product of ${\cal{N}}=4$ SYM matric elements may be more difficult to find. Moreover, even directly, in ${\cal{N}}=8$ supergravity in the unitary gauge,  the clear detailed formulation of this non-linear symmetry still has to be extracted from the fundamental papers  \cite{Cremmer:1978ds} -\cite{de Wit:1982ig} describing the theory.   We will present here the $E_{7(7)}$ symmetry in the unitary $SU(8)$ gauge and the corresponding classically conserved Noether charge of the $E_{7(7)}$ algebra. The linearized version of it with 63 $SU(8)$ symmetries will also be given.

We  will discuss, following \cite{Gunaydin:2000xr},  the  conformal realization of the $E_{7(7)}$ algebra   which has a three-graded  decomposition under the  $E_{6(6)}\times SO(1,1)$ subgroup. One of the 133 charges of  $E_{7(7)}$ algebra is a dilatation operator, which commutes with the generators of $E_{6(6)}$ and gives a positive (negative) weight to the analogs of the translation (special conformal transformations), respectively.
In the context of the  Noether charge constructed above the fact that the conformal realization includes  a dilatation operator may be interesting if $E_{7(7)}$ symmetry has no anomalies in four-dimensional perturbation theory.

Here it is useful to remind that the continuous $E_{7(7)}$ duality symmetry of M/string theory is believed to be broken down to $E_{7(7)}(\mathbb{Z})$ by the existence of extremal four-dimensional black holes when interpreted as massive modes of the fundamental strings as proposed in \cite{Hull:1994ys}, \cite{Witten:1995ex}. In particular, the black hole entropy formula is given by the quartic $E_{7(7)}(\mathbb{Z})$ invariant and depends on the quantized electric and magnetic charges of the black holes \cite{Kallosh:1996uy}. One should also take into account that the compactified to four dimensions string theory may be different from the four-dimensional ${\cal N}=8$ supergravity due to the existence of the non-perturbative states originating from  string theory \cite{Green:2007zzb}.

Thus it is  interesting to find the classically conserved set of Noether charges associated with the classical symmetries ${\cal N}=8$ supergravity and study how quantum corrections respect these symmetries, for example the 63   linear chiral $SU(8)$ symmetries  and the rest of non-linear symmetries of $E_{7(7)}$.

The  paper is organized as follows.
\par
In section 2, we introduce the field content of ${\cal N}=8$ supergravity, including the 56-bein connecting the $E_{7(7)}$ anti-symmetric pair of indices  with the anti-symmetric pair  of $SU(8)$ indices. We present  the structure of global $E_{7(7)}$ and local $SU(8)$ symmetry of the classical action. In section 3, we fix the local $SU(8)$ symmetry by choosing the unitary gauge condition and we  present the gauge fixed action. The theory possesses a generalized electric-magnetic duality symmetry, namely the system of equations of motion for vectors and their corresponding Bianchi identities are covarinat under an $E_{7(7)}$ global transformation. We demonstrate in section 4 how to realize this duality symmetry in the gauge fixed action. We find the explicit exact form of transformations on all fields in a closed form which is valid to all orders of the gravitational coupling constant. In section 4.1, we first explain how this symmetry acts on scalar fields of the theory in the unitary gauge. The transformation rules of vectors and spinors are given respectively in section 4.2 and 4.3.  In section 5, we first embed the $E_{7(7)}$ duality group into the symplectic group $Sp(56,{\mathbb{R}})$ and then we use the results of \cite{Gaillard:1981rj} for theories with electric-magnetic duality to derive the Noether current and its conserved charge associated with the symmetry introduced in section 4. In section 5.1 we derive the linearized form of the Noether current describing the $SU(8)$ symmetry of the asymptotic free fields of ${\cal N}=8$ supergravity.
In section 6  we review the conformal realization of $E_{7(7)}$ algebra.
We discuss the results in sec.  7. Finally, in appendix, we outline the proof of the invariance of the action under the non-linear $E_{7(7)}$ symmetry.

\section{Symmetries of ${\cal N}=8$ Supergravity}
\setcounter{equation}{0}
The field content of ${\cal{N}}=8$ supergravity is: a vierbein field,  $e_\mu{}^a$, 8 gravitino's,  $\psi_\mu{}^i$, 28 abelian gauge fields, ${\cal A}_\mu^{IJ}$, 56 Majorana spinors gaugino, $\chi^{ijk}$. The scalars before the local $SU(8)$ symmetry is fixed belong to a group element of $E_{7(7)}$, so there are 133 scalars. Only 70 of them are physical since one can use the 63 local parameters to bring the theory to the form with 35 complex scalars $\phi_{ijkl}$.
The Lagrangian before gauge-fixing, see (3.18) in \cite{de Wit:1982ig}, consists of 3 parts
\begin{eqnarray}\label{lagrangian}
{\cal L}= {\cal L}_1 + {\cal L}_2+ {\cal L}_3
\end{eqnarray}
The part in ${\cal L}_1$, includes the Einstein term and depends on  fermions. The dependence on scalars is only via the $SU(8)$ connections.    It is vector independent and  manifestly invariant under the linear action of $E_{7(7)}$  as well as under independent local $SU(8)$ symmetry. The part ${\cal L}_2$  has a pure scalar action and part depending on scalars and fermions. Before gauge-fixing of local $SU(8)$ symmetry, it is manifestly invariant under both local $SU(8)$ symmetry as well as under the rigid  $E_{7(7)}$ transformation. Finally,  ${\cal L}_3$ is the part depending on vectors, scalars and fermions.  It is not invariant under $E_{7(7)}$, however, it can be rewritten in the form in which it is invariant under both  $E_{7(7)}$ transformation as well as the local $SU(8)$ symmetry, after partial integration and use of equations of motion,  as explained in \cite{de Wit:1982ig}.

We refer the reader to the details of the  action of ${\cal N}=8$ supergravity and proof of symmetries before the gauge-fixing in \cite{Cremmer:1979up}, \cite{de Wit:1982ig} and will focus here directly on the unitary gauge where only physical 70 scalars are present. We follow notation of \cite{de Wit:1982ig}.

Scalars before gauge-fixing are in the $E_{7(7)}$ group element and there are 133 of them, defined by the 56-bein connecting the $E_{7(7)}$ anti-symmetric pair of indices $IJ$ with the anti-symmetric pair  of $SU(8)$ indices, $ij$.
\begin{eqnarray}\label{gauge}
\cV=\left(
                                        \begin{array}{cc}
                                          u_{ij}{}^{IJ} & v_{ijKL} \\
                                          v^{klIJ} & u^{kl}{}_{KL} \\
                                        \end{array}
                                      \right)\ .
\end{eqnarray}
The rigid  $E_{7(7)}$ symmetry is realized on scalars as a multiplication from the right:
\begin{eqnarray}\label{gauge}
\cV' ({E_{7(7)}}) =\cV E^{-1}
\end{eqnarray}
where $E\in E_{7(7)}$ and is in the fundamental 56-dimensional representation where
\begin{eqnarray}\label{gauge}
  E= \exp G_{E_{7(7)}} =\left(
                                        \begin{array}{cc}
                                        \cA_{IJ}{}^{KL} & \cB _{IJPQ} \\
                                          \cC^{MNKL}&\cD ^{MN}{}_{PQ} \\
                                        \end{array}
                                      \right)
                                       \ , \qquad G_{E_{7(7)}}=\left(
                                        \begin{array}{cc}
                                         \Lambda_{IJ}{}^{KL} & \Sigma _{IJPQ} \\
                                          \Sigma^{MNKL} & \Lambda ^{MN}{}_{PQ} \\
                                        \end{array}
                                      \right)\ .
\end{eqnarray}
The $E_{7(7)}$ Lie algebra requires that
\be
\Lambda_{IJ}{}^{KL}= \delta_{[I}{}^{[K} \Lambda_{J]}{}^{L]}\ .
\ee
In above $\Lambda_{I}{}^{J}$ are the generators of the $SU(8)$ maximal subgroup of $E_{7(7)}$. Therefore, there are 63 of them and they are antihermitian and traceless
\be
\Lambda_I{}^J= - \Lambda ^J{}_I \qquad \Lambda ^I{}_I=0\ .
\ee
They can be decomposed into 28  antisymmetric generators of the $SO(8)$ subgroup and 35 traceless symmetric generators orthogonal to $SO(8)$. If we write $\Lambda_{I}{}^{J}$ as the sum of the real and imaginary parts $\Lambda=\mbox{Re}\Lambda+i\mbox{Im}\Lambda$, then we have
\begin{eqnarray}\label{lambda12}
\mbox{Re}\Lambda^{\mathsf{T}}=-\mbox{Re}\Lambda\ ,\quad \mbox{Im}\Lambda^{\mathsf{T}}=\mbox{Im}\Lambda\ ,
\end{eqnarray}
where the real part is identified with the antisymmetric and the imaginary part with the symmetric part of $\Lambda$.
The off-diagonal part has to satisfy the self-duality constraint with the phase $\eta=\pm1$
\be
\Sigma_{IJKL}= {1\over 24} \eta \, \epsilon_{IJKLMNPQ}  \Sigma ^{MNPQ}\ .
\ee
As the previous case, we can decompose $\Sigma$ into real and imaginary parts $\Sigma=\mbox{Re}\Sigma+i\mbox{Im}\Sigma$. However, in this case, both real and imaginary parts of $\Sigma$ have the same transposition properties
\begin{eqnarray}\label{sigma12}
\mbox{Re}\Sigma^{\mathsf{T}}=\mbox{Re}\Sigma\ ,\quad \mbox{Im}\Sigma^{\mathsf{T}}=\mbox{Im}\Sigma\ .
\end{eqnarray}
Then the self-duality constraint implies that the real part is $\eta$-self-dual and imaginary part is $\eta$-anti-selfdual. The real and imaginary parts of $\Sigma$ each consists of 35 real parameters. Thus we present the 133 real parameters of $E_{7(7)}$  as $133=28+35+35+35$.
\par
The local $SU(8)$ transformation acts on the 56-bein from the left and is completely independent of the $E_{7(7)}$ transformation
\begin{eqnarray}\label{gauge}
\cV' ({SU(8)} ) = U(x) \cV
\end{eqnarray}
where
\begin{eqnarray}\label{gauge}
U(x) = \exp G_{SU(8)} \ ,   \qquad
G_{SU(8)}(x) =\left(
                                        \begin{array}{cc}
                                         \delta_{[i}{}^{[k} \Lambda_{j]}{}^{l]}(x) & 0 \\
                                          0& \delta^{[m}{}_{[p} \Lambda^{n]}{}_{q]}(x)\\
                                        \end{array}
                                      \right)\ .
\end{eqnarray}
Again in order for $\Lambda_{j}{}^{i}$ to be the generators of $\mathfrak{su}(8)$ (Lie algebra associated with $SU(8)$), they must be antihermitian and traceless. They can be decomposed into 28 real antisymmetric and 35 real symmetric parameters.
\section{Gauge Fixing of $SU(8)$ in the Unitary Gauge}
\setcounter{equation}{0}

Before the local $SU(8)$ symmetry is gauge-fixed, there are 133 scalars which form a group element of $E_{7(7)}$.  In this section, we use the local $SU(8)$ symmetry of the action to remove the unphysical 63 scalars from the theory. The local $SU(8)$ symmetry is gauge-fixed in the unitary gauge so that we are left with only 70 dynamical scalar fields.
\par
In the unitary gauge there is no distinction between the $E_{7(7)}$ and $SU(8)$ indices\footnote{\ In ${\mathcal{L}}_{3}$, we still find it convenient to keep capital indices which indicates the transformation under $E_{7(7)}$.}. One can use the 63 local functions $ \Lambda_{j}{}^{l}(x) $ to bring the 56-bein $\cV$ to the form in which the expression in the exponent is vanishing on the diagonal so that the 56-bein matrix becomes hermitian:
\begin{eqnarray}\label{gauge}
\cV=\cV^\dagger\ ,\ \ \ \mbox{where}\ \ \ \cV= \exp  { \left(
                                        \begin{array}{cc}
                                         0&  a\, \phi_{ijkl} \\
                                          a\, \bar \phi ^{mnpq} & 0 \\
                                        \end{array}
                                      \right)} \ .
\end{eqnarray}
where $a= -{\sqrt{2}\over 4}$. Here the  70 physical fields of  ${\cal N}=8$ supergravity are self-dual with the phase $\eta=\pm1$ and completely anti-symmetric in their 4 indices.
\be
\phi_{ijkl}= {1\over 24} \eta \epsilon_{ijklmnpq} \bar \phi^{mnpq}
\ee
The inhomogeneous coordinates of the ${E_{7(7)}\over SU(8)}$ coset space are defined as function of the independent scalars fields $\phi, \bar \phi$ as follows:
\be\label{yphi}
y_{ij, kl}\equiv  \phi_{ijmn}\left({\tanh(\sqrt { {1\over 8} \bar\phi \phi}\over \sqrt {\bar  \phi \phi}} \right)^{mn}_{\; \; \; \;\; {kl}}
\ee
Using the $y$ variables will be very useful in what follows, however, one should keep in mind that they are not independent fields of the theory. The advantage of using these inhomogeneous coordinates of the coset space is in their simple fractional  transformation under the $E_{7(7)}$: this will allow us to derive a
simple closed form  expression for the non-linearly realized symmetry. It will become an
infinite series when expressed in terms of the independent fields of the theory. Note that the dimensionless fields $\phi$ are related to scalars $\varphi$ measured in Planck units as follows
\be
\phi_{ijkl}= \kappa \varphi_{ijkl}
\ee
Thus,  the expansion of $y$ in terms of canonically normalized fields $\varphi, \bar \varphi$ is an expansion in gravitational coupling $\kappa$.
In terms of the nonlinear scalar fields $y, \bar y$,  the matrix $\cV$ is
\begin{eqnarray}\label{Vy}
\cV(y)=\left(
               \begin{array}{cc}
                 P^{-1/2} & -P^{-1/2}y \\
                 -\bar{P}^{-1/2}\bar{y} & \bar{P}^{-1/2} \\
               \end{array}
             \right)\ ,
\end{eqnarray}
where each entry is a $28\times28$ matrix and the matrix $P$ is defined as
\be
P(y,\bar{y})_{ij}{}^{kl}=(\delta_{ij}{}^{kl} -y_{ijrs}\bar{y}^{rskl})
\ee
The action of ${\cal N}=8$ supergravity in the unitary gauge  (\ref{gauge}) consists of 3 terms defined above which can be given, using the $y$-coordinates of the coset space.
\par
The first part of the Lagrangian , ${\mathcal{L}}_{1}$,  which includes the gravitational part, fermions and scalar dependent $SU(8)$ gauge connections,  is given by
\begin{eqnarray}\label{L1}
{\cal L}_{1}&=&-\frac{1}{2}e{\mathcal{R}}(e,\omega)-\frac{1}{2}\epsilon^{\mu\nu\rho\sigma}\big(\bar{\psi}^{i}_{\mu}
\gamma_{\nu}D_{\rho}\psi_{\sigma i}-\bar{\psi}^{i}_{\mu}\overleftarrow{D}_{\rho}\gamma_{\nu}\psi_{\sigma i}\big)
-\frac{1}{12}e\big(\bar{\chi}^{ijk}\gamma^{\mu}D_{\mu}\chi_{ijk}-\bar{\chi}^{ijk}\overleftarrow{D}_{\mu}
\gamma^{\mu}\chi_{ijk}\big)\nonumber\\
&& -\frac{1}{2}e\bar{\psi}_{\mu}^{[i}\psi_{\nu}^{j]}\bar{\psi}^{\mu}_{i}
\psi^{\nu}_{j}+\frac{\sqrt{2}}{4}e\big[\bar{\psi}^{i}_{\lambda}\sigma^{\mu\nu}\gamma^{\lambda}\chi_{ijk}
\bar{\psi}^{j}_{\mu}\psi^{k}_{\nu}+\mbox{h.c.}\big]\nonumber\\
&& +e\big[\frac{1}{144}\eta\varepsilon_{ijklmnrs}\bar{\chi}^{ijk}\sigma^{\mu\nu}\chi^{lmn}\bar{\psi}^{r}_{\mu}
\psi^{s}_{\nu}+\frac{1}{8}\bar{\psi}^{i}_{\lambda}\sigma^{\mu\nu}\gamma^{\lambda}\chi_{ikl}\bar{\psi}_{\mu j}
\gamma_{\nu}\chi^{jkl}+\mbox{h.c.}\big]\nonumber\\
&& +\frac{\sqrt{2}}{864}\eta e\big[\varepsilon^{ijklmnrs}\bar{\chi}_{ijk}\sigma^{\mu\nu}\chi_{lmn}
\bar{\psi}^{p}_{\mu}\gamma_{\nu}\chi_{rsp}+\mbox{h.c.}\big]\nonumber\\
&& +\frac{1}{32}e\bar{\chi}^{ikl}\gamma^{\mu}\chi_{jkl}\bar{\chi}^{jmn}\gamma_{\mu}\chi_{imn}-\frac{1}{96}e
(\bar{\chi}^{ijk}\gamma^{\mu}\chi_{ijk})^{2}\ ,
\end{eqnarray}
Here the covariant derivative acting on fermions
\begin{eqnarray}
&& D_{\mu}\psi_{\nu i}=(\delta_{i}^{j}\partial_{\mu}-\cB^{j}_{\mu i})\psi_{\nu j}\ ,\label{covspin1}\\
&& D_{\mu}\chi_{ijk}=(\delta_{[i}^{l}\partial_{\mu}-3\cB^{l}_{\mu[i})\chi_{jk]l}\ .\label{covspin2}
\end{eqnarray}
 is constructed via the help of the scalar dependent  connection
\begin{eqnarray}\label{Bmu}
\cB^{i}_{\mu j}=\frac{2}{3}\big(\bar{P}^{-1/2}\partial_{\mu}\bar{P}^{-1/2}\big)^{ik}{}_{jk}-\frac{2}{3}\big(\bar{P}^{-1/2}
\bar{y}\partial_{\mu}(y\bar{P}^{-1/2})\big)^{ik}{}_{jk}\ .
\end{eqnarray}
Notice that if we expand (\ref{Bmu}) in terms of $y$ (or alternatively $\phi$), $\cB^{i}_{\mu j}$ only depends on even (odd) powers of scalar fields. The transformation law of this connection is given by
\begin{eqnarray}\label{Bmutrans}
\delta\cB^{i}_{\mu j}=  \Upsilon^{k}{}_{j}\cB^{i}_{\mu k}+\Upsilon_{k}{}^{i}\cB^{k}_{\mu j}+\partial_{\mu}\Upsilon^{i}{}_{j}\ .
\end{eqnarray}
The covariant derivatives   transforms covariantly under the field dependent  $SU(8)$ transformation $\Upsilon$. The explicit form of $\Upsilon$ will be given later.

The second part of the Lagrangian has a kinetic terms for scalars and scalar-fermion interaction term
\begin{eqnarray}\label{L2}
{\cal L}_{2}=-\frac{1}{96}e\ {\cA}_{\mu}^{ijkl}{\cA}^{\mu}_{ijkl} -\frac{1}{24}e\big[\bar{\chi}_{ijk}\gamma^{\nu}\gamma^{\mu}\psi_{\nu l}(\hat{\cA}_{\mu}^{ijkl}+
\cA_{\mu}^{ijkl})+\mbox{h.c.}\big] ,
\end{eqnarray}
where ${\cA}^{ijkl}_{\mu}$ is defined as
\begin{eqnarray}\label{Amu}
{\cA}^{ijkl}_{\mu}=2\sqrt{2}(\frac{1}{\sqrt{1-y\bar{y}}}\partial_{\mu}\bar{y}\frac{1}{\sqrt{1-\bar{y}y}})^{ijkl}\ ,
\end{eqnarray}
where $\hat{\cA}^{ijkl}_{\mu}$ in above is defined as
\begin{eqnarray}\label{hatA}
\hat{\cA}^{ijkl}_{\mu}=\cA^{ijkl}_{\mu}-4\big(\bar{\psi}^{[i}_{\mu}\chi^{jkl]}+\frac{1}{24}\eta
\varepsilon^{ijklmnrs}\bar{\psi}_{\mu m}\chi_{nrs}\big)\ .
\end{eqnarray}
Note that ${\cA}^{ijkl}_{\mu}$ only depends on odd (even) powers of $y$ ($\phi$). The pure scalar part of the action can also be rewritten as
\begin{eqnarray}\label{Lsc}
{\cal L}_{sc}=-\frac{1}{96}e\ {\cA}_{\mu}^{ijkl}{\cA}^{\mu}_{ijkl}=-\frac{1}{12}e\ \mbox{Tr}
\Bigg(\frac{1}{1-y\bar{y}}\partial_{\mu}y\frac{1}{1-\bar{y}y}\partial^{\mu}\bar{y}\Bigg) ,
\end{eqnarray}
Finally, in  ${\cal{L}}_{3}$ we have grouped all terms with vector fields as well as some 4-fermionic terms which are useful for the proof of the symmetry of ${\cal{L}}_{3}$.  All dependence on 28 real Abelian vector fields  ${\cal A}_\mu^{IJ}$ enters only via the $U(1)$ gauge invariant Maxwell field strength
\be
F_{\mu\nu}^{IJ}= \partial_\mu  {\cal A}_\nu^{IJ}- \partial_\nu  {\cal A}_\mu^{IJ}
\ee
which, in turn, can be split into self-dual and anti-self-dual strengths
\be
F_{\mu\nu}^{IJ}= F^{+}_{\mu\nu IJ} +F_{\mu\nu}^{-IJ}  \qquad  , \qquad i \tilde F_{\mu\nu}^{IJ}= F^{+}_{\mu\nu IJ} -F_{\mu\nu}^{-IJ}\ ,
\ee
which are related by complex conjugation $(F^{+}_{\mu\nu IJ})^*= F_{\mu\nu}^{-IJ}$. In above, the dual field strength $\tilde{F}_{\mu\nu}^{IJ}$ is defined as usual as
\be
 i \tilde F_{\mu\nu} \equiv  {1\over 2} \epsilon_{\mu\nu\rho\sigma} F^{\rho\sigma  IJ}\ .
\ee
The self-dual and anti-self dual field strengths are, of course, given by
\be
F^{+}_{\mu\nu IJ}= {1\over 2} (F_{\mu\nu}^{IJ}+  i \tilde  F_{\mu\nu}^{IJ})\qquad,\qquad
F^{-IJ}_{\mu\nu}= {1\over 2} (F_{\mu\nu}^{IJ}-  i \tilde  F_{\mu\nu}^{IJ})\ ,
\ee
in terms of the original field strength $F_{\mu\nu}^{IJ}$ and its dual. The  ${\cal L}_3$ part of the action consists of  terms depending on vectors via $F_{\mu\nu}^{IJ}$ and on fermions \cite{deWit:1982ig}.
\begin{eqnarray}\label{L3}
{\cal{L}}_{3}&=&-\frac{1}{8}e\Big[F^{+}_{\mu\nu IJ}\Bigg(\frac{1+\bar{y}}{1-\bar{y}}\Bigg)^{IJKL}F^{+\mu\nu}_{KL}
+\mbox{h.c.}\Big]\nonumber\\
&&-\frac{1}{2}e\Big[F^{+}_{\mu\nu IJ}\Bigg(\frac{1}{1-\bar{y}}\Bigg)^{IJKL}\O^{+\mu\nu KL}+\mbox{h.c.}\Big]\nonumber\\
&&-\frac{1}{4}e\Big[\O^{+IJ}_{\mu\nu}\big[\Bigg(\frac{1}{1-\bar{y}}\Bigg)^{IJKL}
-\Bigg(\frac{1}{1-y\bar{y}}\ y\Bigg)^{IJKL}\big]\O^{+\mu\nu KL}+\mbox{h.c.}\Big]\ ,
\end{eqnarray}
where the composite fermionic bilinear term is $\O^{+KL}_{\mu\nu}=u^{KL}_{\ \ \ \ ij}\O^{+ij}_{\mu\nu}$, where $\O^{+ij}_{\mu\nu}$ is defined in terms of the original fermionic fields of ${\cal{N}}=8$ supergravity in the following way
\begin{eqnarray}\label{O+}
\O^{+ij}_{\mu\nu}=-\frac{\sqrt{2}}{144}\eta\varepsilon^{ijklmnrs}\bar{\chi}_{klm}
\sigma_{\mu\nu}\chi_{nrs}-\frac{1}{2}\bar{\psi}_{\lambda k}\sigma_{\mu\nu}\gamma^{\lambda}\chi^{ijk}+\frac{\sqrt{2}}{2}\bar{\psi}^{i}_{\rho}\gamma^{[\rho}\sigma_{\mu\nu}
\gamma^{\lambda]}\psi^{j}_{\lambda}\ ,
\end{eqnarray}
and $u^{KL}_{\ \ \ \ ij}= (P^{1/2})^{KL}{}_{ij}$ in the unitary gauge.

  If one fixes the $SU(8)$ local symmetry by the condition $\cV=\cV^\dagger$, one cannot perform an $E_{7(7)}$  transformation anymore since it breaks the gauge-fixing condition. It tends to restore the original 133 scalars by filling in the diagonal terms in the exponent in the equation (\ref{gauge}),  instead of keeping only 70
   off-diagonal physical scalars.
  However,  one can  perform an $E_{7(7)}$ transformation  accompanied by an extra field-dependent $SU(8)$ transformation depending on the rigid parameters of $E_{7(7)}$  symmetry and scalar fields. Only a combination of $E_{7(7)}$  and field dependent $SU(8)$ can preserve the gauge-fixing condition $\cV=\cV^\dagger$. Let us look at it in more detail.

\section{Non-linear Realization of $E_{7(7)}$}
\setcounter{equation}{0}
In this section, we define the transformation laws of scalars, vectors and spinors separately.
The unitary  gauge condition (\ref{gauge}) is preserved if the transformation law includes {\it simultaneous } multiplication of the 56-bein from the right by an arbitrary rigid $E_{7(7)}$ group element depending on 63 rigid parameters  $\Lambda_{J}{}^{L}$ and 70 rigid parameters $\Sigma_{IJKL}$ as well as a multiplication from the left by an compensating $SU(8)$ transformation depending on scalars as well as on 63 rigid parameters  $\Lambda_{J}{}^{L}$ and 70 rigid parameters $\Sigma_{IJKL}$
\begin{eqnarray}\label{Vtrans}
\mathcal{U}(y,\bar{y}; \Lambda, \Sigma) \cV(y')=\cV(y, \bar y) E^{-1}(\Lambda, \Sigma)\ ,
\end{eqnarray}
in which ${\mathcal{U}}\in SU(8)/{\mathbb{Z}_{2}}$ since it acts on $SU(8)$ tensors and $E\in E_{7(7)}$.
Here $\cV(y, \bar y)$ is defined in (\ref{Vy}) and
\begin{eqnarray}\label{Vy'}
\cV(y')=\left(
               \begin{array}{cc}
                 P^{'-1/2} & -P^{'-1/2}y' \\
                 -\bar{P}^{'-1/2}\bar{y} & \bar{P}^{'-1/2} \\
               \end{array}
             \right)\ , \qquad ( P_{ij}{}^{kl})'=\delta_{ij}{}^{kl} -y'_{ijrs}\bar{y}^{'rskl}
\end{eqnarray}
Equivalently
\begin{eqnarray}\label{Vtrans1}
\cV(y')=  \mathcal{U}^{-1}(y,\bar{y}; \Lambda, \Sigma) \cV(y, \bar y) E^{-1}(\Lambda, \Sigma)\ ,
\end{eqnarray}
This is the  definition of scalar variables $y$  and $y'$ before and after the non-linearly realized $E_{7(7)}$ 133-component symmetry transformations. The remaining transformations on vectors and spinors must be consistent with  (\ref{Vtrans1}).

\subsection{Transformation Laws of Scalars}

If we write the above transformation law (\ref{Vtrans})  in its matrix form, we have
\begin{eqnarray}\label{Vmatrix}
\cV(y')=\left(
        \begin{array}{cc}
          \cU^{-1} & 0 \\
          0 & \bar{\cU}^{-1} \\
        \end{array}
      \right)\left(
               \begin{array}{cc}
                 P^{-1/2} & -P^{-1/2}y \\
                 -\bar{P}^{-1/2}\bar{y} & \bar{P}^{-1/2} \\
               \end{array}
             \right)\left(
                      \begin{array}{cc}
                      \cA& -\cB \\
                      -\cC &\cD \\
                      \end{array}
                    \right)\ ,
\end{eqnarray}
which results in the following matrix relations
\begin{eqnarray}\label{mat11-12}
&& P'(y',\bar{y}')^{-1/2}=\cU^{-1}P^{-1/2}(\cA+y\cC)\ ,\\
&& P'(y',\bar{y}')^{-1/2}y'=\cU^{-1}P^{-1/2}(\cB+y\cD)\ .
\end{eqnarray}
Solving these two equations for $y'$ and $\cU$, we obtain
\begin{eqnarray}
&& y'=(\cA+y\cC)^{-1}(\cB+y\cD)\ ,\label{yprime}\\
&& \cU(y, \bar y; \cA, \cB, \cC, \cD)=P(y,\bar{y})^{-1/2}(\cA+y\cC)P'(y',\bar{y}')^{1/2}\ .\label{Usolve}
\end{eqnarray}
Notice that in the last factor of (\ref{Usolve}), the transformed scalar $y'$ is substituted by (\ref{yprime}). As we see, the compensating local $SU(8)$ transformation is not an independent transformation and is correlated with $E_{7(7)}$ transformation. If we consider the infinitesimal $E_{7(7)}$ transformation,
\begin{eqnarray}\label{Einverse}
E^{-1}=\left(
         \begin{array}{cc}
           1+\Lambda & -\Sigma \\
           -\bar{\Sigma} & 1+\bar{\Lambda} \\
         \end{array}
       \right)\ ,
\end{eqnarray}
we can find the explicit forms of $y'$ and $\cU(y, \bar y)$. For $\delta y=y'-y$ , we find
\begin{eqnarray}\label{deltay}
\xymatrix{*+[F]{
\delta y\equiv y'-y=\Sigma+ y\bar{\Lambda}-\Lambda y-y\bar{\Sigma}y\ , }}
\end{eqnarray}
and $\cU(y,\bar{y})$ turns out to be
\begin{eqnarray}\label{Uinfi}
\xymatrix{*+[F]{
\cU(y, \bar y)=1+\frac{1}{2}\ P^{-1/2}\Delta(\Lambda,\Sigma)P^{-1/2}\equiv 1+\Upsilon\ , }}
\end{eqnarray}
where $\Delta(\Lambda,\Sigma)$ is given by
\begin{eqnarray}\label{Delta}
\Delta(\Lambda,\Sigma)=\{\Lambda,P\}+(y\bar{\Sigma}-\Sigma\bar{y})
-y(\bar{\Sigma}y-\bar{y}\Sigma)\bar{y}\ .
\end{eqnarray}

The first terms of the $\kappa$ expansion in the non-linear realization of $E_{7(7)}$ symmetry on scalars and vectors  were  recently presented in \cite{Brink:2008qc}. Here we have found  a relatively simple exact in all orders in $\kappa$ expression for the symmetry transformation in terms of the inhomogeneous coordinates of the ${E_{7(7)}\over SU(8)}$ coset space, the $y$-fields. For example, in (\ref{deltay}) there are terms of 0-order, a shift with the constant parameter $\Sigma$, the  first order terms from the compact generators $\Lambda$ and just terms quadratic in $y$-fields with the parameter $\bar \Sigma$.  This can be translated into an infinite series in the independent scalars $\phi$ using the relation between these fields as shown in (\ref{yphi}). The reason why there is only up to quadratic dependence on $y$ in the infinitesimal transformations is because  the full group transformation on $y$ fields is fractional as shown in (\ref{yprime}). The infinitesimal form of fractional transformation gives only up to quadratic terms, after the denominator in (\ref{yprime}) is expanded with account of the fact that $(\cA+y\cC)^{-1}\approx (1+\Lambda -y\bar \Sigma)^{-1}\approx  1-\Lambda +y\bar \Sigma $ where $\Lambda$ and $\bar \Sigma$ are small.

At the linear level, for example for asymptotic fields, we may be interested in the $SU(8)$ subgroup of the transformations above. In such case $\Sigma=0$ and
\begin{eqnarray}
&&(\delta y)_{lin} = y\bar{\Lambda}-\Lambda y\ ,\label{deltayLin}\\
&&\cU(y, \bar y)_{lin}=1+\Upsilon_{lin}  = 1+\Lambda \ ,\label{UinfiLin} 
\end{eqnarray}
where $\Upsilon_{lin}$ has been substituted by $\Lambda$, using (\ref{Uinfi}) and (\ref{Delta}).

\subsection{Transformation Laws of Vectors}
Following \cite{de Wit:1982ig}, we first define a doublet of field strengths two forms as
\begin{eqnarray}\label{F1F2}
F^{+}_{1IJ}\equiv\frac{1}{2}(G^{+}_{IJ}+F^{+}_{IJ})\ \ \ ,\ \ \
F^{+IJ}_{2}\equiv\frac{1}{2}(G^{+}_{IJ}-F^{+}_{IJ})\ ,
\end{eqnarray}
where dual field strength $G^{+}_{IJ}$ is defined as
\begin{eqnarray}\label{G+}
G^{+}_{IJ}\equiv-\frac{4}{e}\frac{\delta {\cal{L}}_{3}}{\delta F^{+}_{IJ}}\ .
\end{eqnarray}
Using the above definition, we find
\begin{eqnarray}\label{G+y}
G^{+}_{IJ}=\Bigg(\frac{1+\bar{y}}{1-\bar{y}}\Bigg)^{IJKL}F^{+}_{KL}+\Bigg(\frac{2}{1-\bar{y}}
\Bigg)^{IJ}_{KL}\O^{+KL}\ .
\end{eqnarray}
Note that in above, we have suppressed all spacetime indices and it should be considered as a relation between two forms. Now using (\ref{F1F2}), we find
\begin{eqnarray}
&& F^{+}_{1IJ}=\Bigg(\frac{1}{1-\bar{y}}\Bigg)^{IJKL}F^{+}_{KL}+\Bigg(\frac{1}{1-\bar{y}}\Bigg)^{IJ}_{KL}\O^{+KL}\ ,
\label{F1y}\\
&& F^{+IJ}_{2}=\Bigg(\frac{\bar{y}}{1-\bar{y}}\Bigg)^{IJKL}F^{+}_{KL}+\Bigg(\frac{1}{1-\bar{y}}
\Bigg)^{IJ}_{KL}\O^{+KL}\ .\label{F2y}
\end{eqnarray}
On the other hand, we know  how $F^{+}_{1}$ and $F^{+}_{2}$ transform under $E_{7}$
\begin{eqnarray}\label{F1F2E7}
\left(
  \begin{array}{c}
    F^{+}_{1} \\
    F^{+}_{2} \\
  \end{array}
\right)\longrightarrow E\left(
  \begin{array}{c}
    F^{+}_{1} \\
    F^{+}_{2} \\
  \end{array}
\right)\ .
\end{eqnarray}
If we rewrite the above transformation for an infinitesimal

 transformation, we have
\begin{eqnarray}\label{deltaF1F2}
\left(
  \begin{array}{c}
    \delta F^{+}_{1IJ} \\
    \delta F^{+IJ}_{2} \\
  \end{array}
\right)=\left(
          \begin{array}{cc}
            \Lambda_{IJ}^{\mbox{\ \ \ }KL} & \Sigma_{IJKL} \\
            \Sigma^{IJKL} & \Lambda^{IJ}_{\mbox{\ \ \ }KL} \\
          \end{array}
        \right)\left(
                 \begin{array}{c}
                   F^{+}_{1KL} \\
                   F^{+KL}_{2} \\
                 \end{array}
               \right)\ .
\end{eqnarray}
Now, if we use (\ref{F1y}) and (\ref{F2y}), we can find the transformation law of the original field strength $F^{+}$. We obtain
\begin{eqnarray}\label{deltaF+}
\xymatrix{*+[F]{
\delta F^{+}=\delta F^{+}_{1}-\delta F^{+}_{2}\equiv XF^{+}+Y\O^{+}\ ,}}
\end{eqnarray}
where $X$ and $Y$ are defined in the following way
\begin{eqnarray}
&& X=\Big[(\Lambda-\bar{\Lambda}\bar{y})+(\Sigma\bar{y}-\bar{\Sigma})\Big]\frac{1}{1-\bar{y}}\ ,\label{XF}\\
&& Y=\Big[(\Lambda-\bar{\Lambda})+(\Sigma-\bar{\Sigma})\Big]\frac{1}{1-\bar{y}}\ .\label{YF}
\end{eqnarray}
We first notice that both $X$ and $Y$ carry four indices and they can be restored easily by the index structure of (\ref{deltaF+}). The other important thing to notice is that the transformation rule of vectors involves bilinear fermions if and only if $\Lambda$ and $\Sigma$ have imaginary parts. If they are both real, then clearly $Y$ would vanish.
\par
For later purposes, let us find the transformation properties of $G^{+}$ as well. Using (\ref{F1F2}), it is evident that
\begin{eqnarray}\label{deltaG}
\delta G^{+}=\delta F^{+}_{1}+\delta F^{+}_{2}\equiv VF^{+}+W\O^{+}\ ,
\end{eqnarray}
where $V$ and $W$ are found to be
\begin{eqnarray}
&& V=\Big[(\Lambda+\bar{\Lambda}\bar{y})+(\bar{\Sigma}+\Sigma\bar{y})\Big]\frac{1}{1-\bar{y}}\ ,\label{VF}\\
&& W=\Big[(\Lambda+\bar{\Lambda})+(\Sigma+\bar{\Sigma})\Big]\frac{1}{1-\bar{y}}\ .\label{WF}
\end{eqnarray}

If we would be interested only in the linear level transformations on vectors which correspond to the $SU(8)$ subgroup of $E_{7(7)}$ we would get
\begin{eqnarray}\label{deltaF+lin}
(\delta F^{+})_{lin}= \Lambda F^{+}\ , \qquad  \delta F^{-}_{lin}= \bar \Lambda F^{-}\ ,
\end{eqnarray}
In terms of real vector  fields this means that
\begin{eqnarray}\label{deltaF+lin}
(\delta F)_{lin} =\mbox{Re}\Lambda \,  F -  \mbox{Im }\Lambda \, \tilde F
\end{eqnarray}
The linearized $SU(8)$ transformation on spinors has the following property: the 28 $SO(8)$ subgroup transformations rotate the real $F$ into itself whereas the 35 orthogonal transform $F$ into the dual field strength $\tilde F$. This agrees with the fact that the positive (negative) helicity states transform by all 63 generators of the $SU(8)$ into itself
\begin{eqnarray}\label{deltaF+lin}
(\delta F^{\pm})_{lin}= (\mbox{Re}\Lambda \pm i \mbox{Im }\Lambda ) F^{\pm}\ .
\end{eqnarray}
This explains the meaning of ``chiral'' $SU(8)$.

\subsection{Transformation Laws of Spinors}

Now, in this section, we have to introduce the transformations on spinor fields in such a way that the Lagrangian remains invariant under simultaneous performance of (\ref{deltay}), (\ref{Uinfi}), and (\ref{deltaF+}). If the spinors transform covariantly by the compensating local $SU(8)$ transformation as $SU(8)$-tensors, then every term in the Lagrangian which involves fermions remains invariant manifestly, because it will be an $SU(8)$-scalar. The proof of the invariance of all different pieces of Lagrangian will be discussed in the appendix in more details.
\par
If we rewrite (\ref{Uinfi}) as $\cU_{ij}{}^{kl}\equiv\delta_{[i}{}^{[k}\Theta_{j]}{}^{l]}$, then $\Theta_{i}{}^{j}$ is given by the partial trace of $\cU$ as
\begin{eqnarray}\label{Theta}
\Theta_{i}{}^{l}(\Lambda,\Sigma)=\delta_{i}{}^{l}+\Upsilon_{ij}{}^{lj}(\Lambda,\Sigma) \equiv \delta_{i}{}^{l}+\Upsilon_{i}{}^{l}(\Lambda,\Sigma) \ ,
\end{eqnarray}
where $\Upsilon_{i}{}^{l}(\Lambda,\Sigma)$ is explicitly given by
\begin{eqnarray}\label{upsilontrace}
\Upsilon_{i}{}^{l}(\Lambda,\Sigma)=\frac{1}{2}(P^{-1/2})_{ij}{}^{mn}\Delta_{mn}{}^{rs}(\Lambda,\Sigma)
(P^{-1/2})_{rs}{}^{lj}\ .
\end{eqnarray}
Then the transformation law of the left-handed gravitini are given by
\begin{eqnarray}\label{gravitinotran}
\xymatrix{*+[F]{
\delta\psi_{i \mu}=\Upsilon_{i}{}^{j}\psi_{j \mu}\ ,}}
\end{eqnarray}
and for gaugini, we have
\begin{eqnarray}\label{gauginotran}
\xymatrix{*+[F]{
\delta\chi_{ijk}=3 \Upsilon_{[i}{}^{q}\chi_{jk]q }\ ,}}
\end{eqnarray}
respectively\footnote{\ Note that for finite values of $\Lambda$ and $\Sigma$, the transformation law of gaugini will be
\begin{eqnarray}
\chi'_{ijk}=\Theta_{[i}{}^{q}\Theta_{j}{}^{r}\Theta_{k]}{}^{s}\chi_{qrs}\ .
\end{eqnarray}
But when we want to find the infinitesimal transformations, we drop all quadratic and higher order terms in $\Lambda$ and $\Sigma$ and that is why we get a factor of 3 in (\ref{gauginotran}).}. This is consistent with the definition of spinors in notation of \cite{deWit:1977fk} which we are using here, namely
\be
 \psi_{i \mu}={ 1-\gamma_5\over 2 }\psi_{i \mu} \, \qquad \chi_{ ijk}= { 1-\gamma_5\over 2 } \chi_{ ijk}
 \ee
etc.
Notice that not only spinor fields, but also their covariant derivatives (\ref{covspin1}) and (\ref{covspin2}) transform as $SU(8)$-tensors via the $SU(8)$-connection (\ref{Bmu}).

We will also need the transformation rules of $\bar{\psi}^{i}_{\mu}$ and $\bar \chi^{ijk}$ which are given by
\begin{eqnarray}
&& \delta \bar \psi^i _{ \mu}=\bar \psi^j _{ \mu}  \Upsilon^{i}{}_{j}\ ,\label{gravitinotranup}\\
&& \delta\bar \chi^{ijk}=3\bar \chi^{q[ij} \Upsilon^{k]}{}_{q}\ .\label{gauginotranup}
\end{eqnarray}
Notice that $\Upsilon$ is an element of ${\mathfrak{su}}(8)$ and therefore it is anti-hermtian $\Upsilon_{j}{}^{i}+\Upsilon^{i}{}_{j}=0$.

\section{Conserved Noether Current}
\setcounter{equation}{0}
The expression for the conserved Noether current for any supergravity theory with duality symmetry of the type discussed in this paper was derived in \cite{Gaillard:1981rj}. All we have to do here is to specialize to the explicit form of the $E_{7(7)}$ symmetry which we have presented above. For this purpose it is convenient to embedd our construction into a symplectic basis of the $Sp(56,{\mathbb{R}})$ matrix, where we have 2 different real field strengthes $(F, G)$ (not independent) which form the fundamental representation of the $E_{7(7)}$
\begin{eqnarray}\label{symplectic}
\left(
                      \begin{array}{cc}
                  F' \\
                    G' \\
                      \end{array}
                    \right)\ ={\cal S}  \left(
                      \begin{array}{cc}
                  F \\
                    G \\
                      \end{array}
                    \right) \ ,  \qquad  {\cal S} \equiv \left(
                      \begin{array}{cc}
                      A& B \\
                       C &D \\
                      \end{array}
                    \right)\ .
\end{eqnarray}
Here the parameters in the symplectic matrix ${\mathcal{S}}$ are real and satisfy the following conditions
\begin{eqnarray}\label{symprule}
A^{\mathsf{T}} C-C^{\mathsf{T}} A= B^{\mathsf{T}}D- D^{\mathsf{T}}B=0\ , \qquad A^{\mathsf{T}}D- C^{\mathsf{T}}B=1\ .
\end{eqnarray}
This representation was extremely useful in the studies of extremal black holes in extended supergravities where the role of $E_{7(7)}$ was very important and was crucial for the defining the black hole entropy formula \cite{Kallosh:1996uy} via the quartic invariant of $E_{7(7)}$.
\par
Now, we need to find the explicit form of symplectic embedding of $E_{7(7)}$. Substituting (\ref{G+y}) for (\ref{deltaF+}), we find
\begin{eqnarray}\label{deltaF+1}
\delta F^{+}=\Big(\mbox{Re}\Lambda-\mbox{Re}\Sigma\Big)F^{+}+i\Big(\mbox{Im}\Lambda+\mbox{Im}\Sigma\Big)G^{+}\ .
\end{eqnarray}
If we do the same thing for $G^{+}$ by substituting (\ref{G+y}) in (\ref{deltaG}), we can express its variation in terms of $F^{+}$ and $G^{+}$. We find
\begin{eqnarray}\label{deltaG+1}
\delta G^{+}=i\Big(\mbox{Im}\Lambda-\mbox{Im}\Sigma\Big)F^{+}+\Big(\mbox{Re}\Lambda+\mbox{Re}\Sigma\Big)G^{+}\ .
\end{eqnarray}
Now, it is easy to find transformation laws of the actual real field strengths
\begin{eqnarray}\label{realFG}
F^{IJ}=F^{+}_{IJ}+F^{-IJ}\ \ \ ,\ \ \ G^{IJ}=i(G^{+}_{IJ}-G^{-IJ})\ .
\end{eqnarray}
Therefore, under duality transformations, $F$ and $G$ transform in the following way
\begin{eqnarray}\label{deltaFG}
\left(
  \begin{array}{c}
    \delta F \\
    \delta G \\
  \end{array}
\right)=\left(
          \begin{array}{cc}
             \mbox{Re}\Lambda-\mbox{Re}\Sigma & \mbox{Im}\Lambda+\mbox{Im}\Sigma \\
            -\mbox{Im}\Lambda+\mbox{Im}\Sigma & \mbox{Re}\Lambda+\mbox{Re}\Sigma \\
          \end{array}
        \right)=\left(
                  \begin{array}{c}
                    F \\
                    G \\
                  \end{array}
                \right)\ .
\end{eqnarray}
Comparing the above relation with (\ref{symplectic}), we should identify the elements of the symplectic matrix ${\mathcal{S}}$ as the following
\begin{eqnarray}
&& A=1+\mbox{Re}\Lambda-\mbox{Re}\Sigma\ ,\quad B=\mbox{Im}\Lambda+\mbox{Im}\Sigma\ ,\label{sympA}\\
&& C=-\mbox{Im}\Lambda+\mbox{Im}\Sigma\ ,\quad D=1+\mbox{Re}\Lambda+\mbox{Re}\Sigma\ .\label{sympC}
\end{eqnarray}
The only thing which remains to be proved is that the symplectic rules (\ref{symprule}) are indeed satisfied. This can easily be verified using (\ref{lambda12}), (\ref{sigma12}) and the fact that $\Lambda$ and $\Sigma$ are infinitesimal parameters of transformation.
\par
The kinetic terms for the vectors is such that
\begin{eqnarray}\label{G}
\tilde G \equiv 4\frac{\delta {\cal{L}}_{3}}{\delta F}\ .
\end{eqnarray}
When equations of motion and Bianchi identities are satisfied
\be
\partial_\mu \tilde F^{\mu\nu}= \partial_\mu \tilde G^{\mu\nu}=0\ ,
\ee
both $F$ and $G$ can be represented as the derivative of the vector potentials
\bea
\left(
                      \begin{array}{cc}
                  F_{\mu\nu} \\
                    G_{\mu\nu} \\
                      \end{array}
                    \right)\ =
\left(
                      \begin{array}{cc}
                \partial_\mu  {\cal A}_\nu -  \partial_\nu  {\cal A}_\mu \\
               \partial_\mu  {\cal B}_\nu -  \partial_\nu  {\cal B}_\mu \\
                      \end{array}
                    \right)\ .
\eea
Noether current was computed in \cite{Gaillard:1981rj} for a general class of supergravities with duality symmetry given by (\ref{symplectic}) for vector fields and by some transformations of scalars and fermions $\Phi$ of the type
\be
\delta \Phi^a= \xi^a (\Phi; A, B, C, D)\ ,
\ee
where $ \xi^a (\Phi; A, B, C, D)$ is some non-derivative function of fields $\Phi$ and parameters of $E_{7(7)}$ symmetry.
In our case $\Phi^a$ includes all scalars and fermions of ${\cal N}=8$ supergravity
\be
\Phi^a\in \{ \phi_{ijkl}, \chi^{ijk}, \psi_\mu^i \}\ .
\ee
The conserved Noether current consists of 2 parts
\be
J^{\mu}_{total} = \hat J^\mu + J^\mu (\Phi) \ ,  \qquad \partial_\mu J^\mu_{total} =0\ .
\ee
The standard part related to all fields $\Phi^a$ is
\be
J^\mu (\Phi)= \xi^a {\partial {\cal L} \over \partial \Phi_{,\mu}}\ .
\ee
The explicit form of this current can be deduced from the action by replacing every occurrence of the derivative $\partial_\mu$ acting on the scalar, gaugino and gravitino by their corresponding  $E_{7(7)}$ transformation. The transformation on scalars is given in (\ref{deltay}) and on fermions we have to perform an $SU(8)$ transformation with the scalar dependent parameter given in (\ref{Uinfi}).

The Gaillard-Zumino part of the   Noether current $\hat J^\mu$ which depends on the vectors of the theory is unusual,  and is given by
\be \label{vec}
\hat J^\mu({\cal A}, {\cal B})= {1\over 4}(\tilde G^{\mu\nu} (A-1) \, {\cal A}_\nu -\tilde F^{\mu\nu} C \, {\cal A}_\nu +\tilde G^{\mu\nu} B  \, {\cal B}_\nu - \tilde F^{\mu\nu} (D-1) \, {\cal B}_\nu)\ .
\ee
Notice that (\ref{vec}) is the Noether contracted current which means that the Noether current has been contracted with 133 parameters of transformation which are functions of $\Lambda$ and $\Sigma$. Here the symplectic parameters $A, B, D, D$ are functions of $\Lambda$ and $\Sigma$ as shown in eq. (\ref{sympA}).

The vector current  has a form of the combination of Chern-Simons terms such that its derivative is proportional to
\be \label{CST}
\partial_\mu \hat J^\mu({\cal A}, {\cal B})= {1\over 8}(\tilde G^{\mu\nu} (A-1) \, F_{\mu\nu} -\tilde F^{\mu\nu} C \, F^{\mu\nu} +\tilde G^{\mu\nu} B  \, G_{\mu\nu} - \tilde F^{\mu\nu} (D-1) \, G_{\mu\nu})\ .
\ee
There are 4 types of  standard Noether current contributions  from gravitino, from gaugino,  from pure scalars and from the term where fermion bilinears are mixed with the scalar derivative.
\begin{eqnarray}
&&J^\mu (\psi)= -{1\over 8} \epsilon^{\rho\nu\mu\sigma}( \bar \psi_\rho^i \gamma_\nu (1-\gamma_5)\delta\psi_{\sigma i} - \delta \bar \psi_\rho^i \gamma_\nu (1-\gamma_5)\psi_{\sigma i})\ ,\label{gravitino}\\
&& J^\mu (\chi)= -{1\over 24} ( \bar \chi^{ijk} \gamma^\mu (1-\gamma_5) \delta\chi_{ ijk}-  \delta \bar \chi^{i jk} \gamma^\mu (1-\gamma_5)\chi_{ ijk})\ ,\label{gaugino}\\
&& J^\mu (y)= -{1\over 12 } (\bar{P}^{-1/2}\delta \bar y P^{-1})
\partial_\mu  y  \bar P^{-1/2}) +\bar{P}^{-1/2}\partial_\mu \bar y  P^{-1}
\delta  y  \bar P^{-1/2}\ ,\label{scalar}\\
&& J^\mu (y, \chi, \psi)= {\sqrt 2 \over 12 } \bar \chi_{ijk}(1-\gamma_5) \gamma^\nu \gamma^\mu \psi_{\nu l} (\bar{P}^{-1/2}\delta \bar y P^{-1/2})^{ijkl} +\mbox{h.c.}\ .\label{mix}
\end{eqnarray}
Here the expressions for the field variations are given in (\ref{gravitinotran}), (\ref{gauginotran}), and (\ref{deltay}). We have inserted the $\gamma^5$ dependence into the expression of the currents to stress the chiral nature of the symmetries.

The total conserved Noether current is given by the sum of vector and other fields contributions in eqs. (\ref{vec})-(\ref{mix}).
\be
J^{\mu}_{total} = \hat J^\mu ({\cal A}, {\cal B}) + J^\mu (\Phi) \ ,  \qquad \partial_\mu J^\mu_{total} =0\ ,
\ee
where
\be
J_\mu(\Phi)\equiv J^\mu (\psi)+ J^\mu (\chi)+ J^\mu (y)+ J^\mu (y, \chi, \psi)\ .
\ee
Without the Chern-Simons type vector field contribution the current $J_\mu(\Phi)$ is gauge-invariant but not conserved.
\be
\partial_\mu J^\mu (\Phi)= - {1\over 8}(\tilde G^{\mu\nu} (A-1) \, F_{\mu\nu} -\tilde F^{\mu\nu} C \, F^{\mu\nu} +\tilde G^{\mu\nu} B  \, G_{\mu\nu} - \tilde F^{\mu\nu} (D-1) \, G_{\mu\nu})\ .
\ee
It has been noticed in \cite{Gaillard:1981rj} that the total current in (\ref{vec}) is not $U(1)$ gauge invariant since it depends on the Chern-Simons type terms. Under the corresponding $U(1)$ local transformations the total current transforms via a divergence of the antisymmetric tensor
\be
 J^\mu_{total} \rightarrow  J^\mu_{total}  + \partial_\nu X^{[\mu\nu]}\ .
\ee
The corresponding Noether charge
\be
Q_{Noether}=  \int d^3x J^0_{total}\ ,
\ee
is therefore time independent and $U(1)$ gauge invariant.
One can try to study the matrix elements of this conserved current sandwiched  between some initial and final physical states and check the constraints which the conservation of the current imposes.

\subsection{ A Linear $SU(8)$ Part of the $E_{7(7)}$ Noether Charge}

If we restrict ourselves only to $\Lambda$ part of the transformations and only keep the linear parts of transformations, we can extract here the linear part of the $SU(8)$ Noether charge which, for example, can act on asymptotic states. In this manner, the linear part of the Noether charge can easily be applied to study the properties of the S-matrix elements under this symmetry.
\par
In the currents we will keep only quadratic in fields terms. We have to require that $\Sigma=0$ for the consistency of the linear approximation since the shift term in the scalar transformation  is zero order in the fields.  Thus we can present the corresponding linearized $SU(8)$ Noether charge if we keep only linear terms in symmetries,  as well as in all fields and quadratic in fields currents. This gives us
\be
j^{\mu}_{total} = \hat j^\mu ({\cal A}, {\cal B}) + j^\mu (\psi)+ J^\mu (\chi)+ j^\mu (y)+ j^\mu (y, \chi, \psi) \ ,  \qquad \partial_\mu j^\mu_{total} =0\ ,
\ee
where
\be \label{vecLin}
\hat j^\mu({\cal A}, {\cal B})= {1\over 4}(\tilde G^{\mu\nu}   \mbox{Re}\Lambda \, {\cal A}_\nu +\tilde F^{\mu\nu} \mbox{Im}\Lambda \, {\cal A}_\nu +\tilde G^{\mu\nu} \mbox{Im}\Lambda  \, {\cal B}_\nu - \tilde F^{\mu\nu}   \mbox{Re}\Lambda \, {\cal B}_\nu)\ ,
\ee
or
\be
\partial_\mu \hat j^\mu= {1\over 8}(\tilde G   \mbox{Re}\Lambda \, F +\tilde F\mbox{Im}\Lambda \, F +\tilde G \mbox{Im}\Lambda  \, G - \tilde F   \mbox{Re}\Lambda \, G)\ .
\ee
Here we have to use the linear relation between $\tilde G$ and $F$.    It is $F=-\tilde G$  and $G=\tilde F$ \footnote{Note that with the Lorentzian signature, the Hodge star has the following property $(*)^{2}=-1$.}.  In such case we get
\be \label{vecLinSimple}
\hat j^\mu({\cal A}, {\cal B})= {1\over 4}(- F^{\mu\nu}   \mbox{Re}\Lambda \, {\cal A}_\nu +\tilde F^{\mu\nu} \mbox{Im}\Lambda \, {\cal A}_\nu - F^{\mu\nu} \mbox{Im}\Lambda  \, {\cal B}_\nu - \tilde F^{\mu\nu}   \mbox{Re}\Lambda \, {\cal B}_\nu)\ ,
\ee
and
\be
\partial_\mu \hat j^\mu({\cal A}, {\cal B})={1\over 8}(-F  \mbox{Re}\Lambda \, F +\tilde F\mbox{Im}\Lambda \, F -F \mbox{Im}\Lambda \tilde F - \tilde F   \mbox{Re}\Lambda \, \tilde F)=0\ .
\ee
The first and the last term vanish each
\be
F  \mbox{Re}\Lambda \, F =  \tilde F   \mbox{Re}\Lambda \, \tilde F=0\ ,
\ee
since for the $SO(8)$ part of the symmetry, we have $\mbox{Re}\Lambda=-\mbox{Re}\Lambda^{\mathsf{T}}$. The second and the third term cancel each other since the $SU(8)$ compliment to $SO(8)$ has the property $\mbox{Im}\Lambda =\mbox{Im}\Lambda ^{\mathsf{T}}$
\be
\tilde F\mbox{Im}\Lambda \, F -F \mbox{Im}\Lambda \tilde F =0\ .
\ee
The other linearized currents are
\begin{eqnarray}
&& j^\mu (\psi)= -{1\over 4} \epsilon^{\rho\nu\mu\sigma}( \bar {\psi}_{\rho}^{i} \gamma_\nu (1-\gamma_5)\Lambda_{i}{}^{j} \psi_{\sigma j} -   \bar {\psi}_{\rho}^{i} \Lambda^{j}{}_{i} \gamma_\nu (1-\gamma_5)\psi_{\sigma j})\ ,\label{gravitinocurlin}\\
&&j^\mu (\chi)= -{1\over 8} ( \bar \chi^{ijk} \gamma^\mu (1-\gamma_5) \Lambda_{[i}{}^{r} \chi_{jk]r} -   \bar \chi^{r[jk} \Lambda^{i]}{}_{r} \gamma^\mu (1-\gamma_5)\chi_{ijk})\ ,\label{gauginocurlin}\\
&& j^\mu (\phi)= -{1\over 12 } ( \bar \phi \Lambda - \bar \Lambda \bar \phi)
\partial_\mu  \phi   +\partial_\mu \bar \phi  ( \phi  \bar \Lambda- \Lambda \phi)
\ .\label{scalarcurlin}
\end{eqnarray}
We have explicitly presented the indices of an $SU(8)$ generator $\Lambda$ in the first two currents. In the scalar part of the current (\ref{scalarcurlin}), $\Lambda$ acts on $\phi$ in the same way as it acts on $y$. For instance, for the first term $(\bar{\phi}\Lambda)^{ijkl}=\bar{\phi}^{ijmn}\Lambda_{mn}{}^{kl}$. The mixed current  $J^\mu (y, \chi, \psi)_{lin}$ is cubic in fields,  therefore it drops from the linear level approximation, relevant for asymptotic states.

Thus classically the divergence of the vector and pseudo-vector contribution from  gravitino, gaugino and  scalars  vanishes. Therefore, at the classical level the total contribution to the 63-component divergence of the Noether current from all fields  is vanishing, the contribution from vector fields vanishes separately.

\section {On Conformal Realization of $E_{7(7)}$ Algebra}

The purpose of this section is to present the known conformal realization of the $E_{7(7)}$ algebra which has a three-graded  decomposition under the  $E_{6(6)}\times SO(1,1)$ subgroup  \cite{Gunaydin:2000xr}. The fact that the conformal realization includes  a dilatation operator may be interesting, if there are no anomalies associated with $E_{7(7)}$ symmetry in the four-dimensional perturbative theory.

The three-graded  decomposition
\be
\mathfrak{g}= \mathfrak{g}^{-1} \oplus  \mathfrak{g}^0 {} \oplus \mathfrak{g}^{+1}\ ,
\ee
of  $E_{7(7)}$  \cite{Gunaydin:2000xr} consists of splitting all generators of the algebra into
\bea
\bf {133= 27 \oplus (78\oplus 1) \oplus \overline {27}}\ .
\eea
The ${\bf 78}= {\bf 36}+ {\bf 42}$ generators of  the $E_{6(6)}$ algebra consist of the $ {\bf 36}$,   in the adjoint representation of $USp(8)$, symmetric $\bf G^{ij}$ and  the $ {\bf 42}$ in a fully antisymmetric traceless representation  of $USp(8)$, $\bf G^{ijkl}$. It is traceless with respect to the real symplectic metric. Together with the singlet $\bf H$, the generator of $SO(1,1)$,  these 79 generators belong to the $\mathfrak{g}^0$ part of the algebra.The ${\bf 27}$ belong to $\mathfrak{g}^{-1}$ (the analog of translation), and ${\bf \overline 27}$ belong to $\mathfrak{g}^{+1}$ (the analog of special conformal transformations). They are given by fundamental in $E_{6(6)}$ antisymmetric traceless generators $\bf E^{ij}$ and $\bf F^{ij}$,  respectively.

The ${\bf 78}$ generators form the  $E_{6(6)}$ algebra and commute with $\bf H$. The commutator of $\bf H$ with ${\bf 27}$ gives $-{\bf 27}$ and the commutator of $\bf H$  with ${\bf \overline {27}}$ gives $+{\bf \overline {27}}$.

The decomposition of $E_{7(7)}$ algebra  in terms of fields of ${\cal N}=8$ supergravity in four dimensions is natural via the $SU(8)$ subgroup. However, it may be useful also to look at the $E_{6(6)}\times SO(1,1)$ decomposition of the Noether charges of $E_{7(7)}$ which can be deduced from the Kaluza-Klein reduction of the ${\cal N}=8$ supergravity in five dimensions. These two approaches are complimentary.

\section{Discussion}

The 133 parameter $E_{7(7)}$ symmetry of the on-shell action of ${\cal N}=8$ supergravity in the unitary gauge is described in this paper in details.  We have presented  a relatively simple closed  form of the transformations acting on all fields of the theory in the unitary gauge using the inhomogeneous coordinates of the coset space ${E_{7(7)}\over SU(8)}$.

The associated conserved Noether current has 133 components, as expected,  63 of them correspond to compact generators of $E_{7(7)}$ and 70 to the non-compact ones. We have presented here this  exact 133-component Noether current which is conserved when the  classical non-linear equations of motion are satisfied. We also presented a  linearized  $SU(8)$ part of the Noether current and charge.

It is known that the superspace on-shell geometry is described by the torsion and curvature of the  superspace \cite{Brink:1979nt}. These geometric objects are invariant under $E_{7(7)}$ transformations and covariant under $SU(8)$ symmetry. The candidates for the UV divergences starting from the 8-loop order are completely geometric, they depend on torsion and curvature in superspace and therefore they are invariant both under $E_{7(7)}$ and $SU(8)$ symmetry \cite{Howe:1980th},  \cite{Kallosh:1980fi}. From this perspective it is not  clear at present what can be the underlying reason for   ${\cal N}=8$ supergravity to be finite to all loop order, the possibility of which was  proposed in \cite{Bern:2006kd}.

To clarify the situation it may be helpful to understand better the constraints from the non-linearly realized $E_{7(7)}$ symmetry in the unitary gauge with the fixed $SU(8)$ local symmetry. We have presented the detailed description of this symmetry in this paper by decomposing it into the $SU(8)$ subgroup and the orthogonal complement to it, in agreement with the structure of the four-dimensional fields of ${\cal N}=8$ supergravity.
We have also  noticed that an alternative realization of these 133 Noether charges may be useful due to $E_{6(6)}\times SO(1,1)$ decomposition of the symmetry algebra which would relate the symmetry to the version of the theory in terms of the five-dimensional fields of ${\cal N}=8$ supergravity and Kaluza-Klein modes.

It would be interesting to check how  the $E_{7(7)}$ and its linearly realized chiral $SU(8)$ subgroup restrict the
S-matrix of the theory, preferably in the form of the helicity amplitudes used in the computations in \cite{Bern:2007hh}, \cite{Bern:2006kd}. And the important issue remains to see if any of these symmetries are free of anomalies and whether any of this may be useful for understanding of  the UV properties of ${\cal N}=8$ supergravity in four dimensions.

\section*{Acknowledgments}

We are grateful to Tom Banks, Lance Dixon and  Dan Freedman  for the most stimulating and fruitful discussions. We are also thankful to Dan Freedman for sharing with us the results of his work \cite{BEF}. We would also like to thank T. Kugo for pointing out the typos of the first version of this paper. This work is supported by the NSF grant 0244728.

\appendix
\section{Proof of the Gauge-fixed Action Invariance under  Non-linearly Realized $E_{7(7)}$}
\setcounter{equation}{0}
Here, we show the main steps to prove the invariance of different pieces of Lagrangian under simultaneous performance of scalar transformation (\ref{deltay}), compensating local $SU(8)$ transformation (\ref{Uinfi}), and their corresponding vector (\ref{deltaF+}) and spinor transformations (\ref{gravitinotran}) and (\ref{gauginotran}).
\subsection{The ${\mathcal{L}}_{1}$ Part}

The proof of the invariance of the gauge-fixed action given in eqs. (\ref{L1}), (\ref{L2}),  (\ref{L3}) is closely related to the one before gauge-fixing in \cite{Cremmer:1979up}, \cite{de Wit:1982ig},  where the parameters of the local $SU(8)$ are arbitrary 63 local functions, whereas in the gauge-fixed case they are functions of fields and rigid 133 parameters of $E_{7(7)}$. We bring up here the outline of the proof for the convenience of the reader and refer for details to the original papers  \cite{Cremmer:1979up}, \cite{de Wit:1982ig}.

The key point is that  the $SU(8)$ connection  transforms in  the same way as before and after the gauge-fixing. Namely the connection $\cB^{i}_{\mu j}$ in (\ref{Bmu}) depending on $y, \bar y$ transforms as shown in  (\ref{Bmutrans}), which means that the inhomogeneous term $\partial_{\mu}\Upsilon^{i}{}_{j}$  as well as two other homogeneous  terms $\Upsilon^{k}{}_{j}\cB^{i}_{\mu k}+\Upsilon_{k}{}^{i}\cB^{k}_{\mu j}$ are all present. The difference with non-gauge-fixed case is that   an  arbitrary anti-hermitian $x$-dependent matrix $\Lambda^{i}{}_{j}(x)$ has to be replaced by $\Upsilon^{i}{}_{j}(y, \bar y; \Lambda, \Sigma)$ defined in (\ref{Uinfi}).

In the ${\cal L}_1$ part of the action given in (\ref{L1})  each term before $SU(8)$ gauge-fixing is an invariant product of local SU(8) tensors.  The metric does not transform under $SU(8)$ and all fermions are either in $\mathbf{8} $, $\mathbf{\bar 8} $ or in $\mathbf{56} $, $\mathbf{\bar 56} $. The derivatives on spinors are $SU(8)$-covariant and they remain such after the local $SU(8)$ is fixed. Therefore this part of the action remains invariant when the spinors transform under the field-dependent compensating $SU(8)$ transformation together with the relevant transformation of scalars which preserves the correct transformations of the scalar-dependent $SU(8)$ gauge connections in covariant derivatives.

\subsection{The ${\mathcal{L}}_{2}$ Part}

In $ {\cal L}_2$ part of the action, in (\ref{L1}), we want to prove that the pure scalar part ${\cal L}_{sc}$ of the nonlinear sigma model action is invariant under the symmetry (\ref{Vtrans}) in the unitary gauge as well as the scalar-fermion part.   Using (\ref{Vy}), one finds
\begin{eqnarray}\label{dV-V-1}
(\partial_{\mu}\cV)\cV^{-1}=-\frac{\sqrt{2}}{4}\left(
                              \begin{array}{cc}
                                \cB^{i}_{\mu j} \delta^k{}_ l& \cA_{\mu}^{ijkl} \\
                                \cA_{\mu ijkl} & \cB_{\mu j}{}^{i}  \delta_l{}^k \\
                              \end{array}
                            \right)\ ,
\end{eqnarray}
where $A_{\mu}^{ijkl}$ is completely antisymmetric and self-dual and is given by (\ref{Amu}). Also, $\cB^{i}_{\mu j}$ is given in (\ref{Bmu}). Notice that both $A_{\mu}^{ijkl}$ and $\cB^{i}_{\mu j}$ are invariant under global $E_{7(7)}$ transformations since $\cV$ is multiplied by $E^{-1}$ from the right and $\cV^{-1}$ is multiplied on $E$ from the left.
 However, under local $SU(8)$ transformation, $\cA_{\mu}^{ijkl}$ transforms covariantly, whereas $\cB^{i}_{\mu j}$ behaves as an $SU(8)$ connection.

 Therefore, one can  define an $SU(8)$  covariant derivative as follows
\begin{eqnarray}\label{covder2}
(D_{\mu}\cV)\cV^{-1}=-\frac{\sqrt{2}}{4}\left(
                                          \begin{array}{cc}
                                            0 & \cA_{\mu}^{ijkl} \\
                                            \cA_{\mu ijkl} & 0 \\
                                          \end{array}
                                        \right)=\cP^{\mu}\ .
\end{eqnarray}
This shows explicitly that  $\cP_{\mu}$ belongs to the orthogonal complement of $\mathfrak{su}(8)$. The pure scalar Lagrangian in terms of  $y_{ijkl}$ scalars has the following form
\begin{eqnarray}\label{lag2}
{\mathcal{L}}=-\frac{1}{2}\mbox{Tr}\Big((D_{\mu}{\cV})\cV^{-1} (D^{\mu}\cV)\cV^{-1}\Big) = -\frac{1}{2}\mbox{Tr}\big(\cP^{\mu}\cP_{\mu}\big)\ .
\end{eqnarray}
It is manifestly invariant under the transformations
\begin{eqnarray}\label{non-linear}
\cV\longrightarrow \cV'= {\mathcal{U}}(y,\bar{y};\Lambda, \Sigma) \cV E^{-1}(\Lambda, \Sigma) \ .
\end{eqnarray}
Another  term in $ {\cal L}_2$ is a product of an $SU(8)$ tensor $A_{\mu}^{ijkl}$ and fermion bilinears which are also tensors in $SU(8)$ and neutral in $E_{7(7)}$. There is also a terms quartic in fermions which is manifestly invariant. So the total $ {\cal L}_2$ is invariant under (\ref{non-linear}).
\par
Of course, the proof of invariance can be seen in a more general setting by construction. Let us, mention briefly this set up. What we want is to prove that the pure scalar part of the nonlinear sigma model action is invariant under the symmetry (\ref{Vtrans}) in the unitary gauge. This is shown by construction, namely we write down a Lagrangian in terms of 70 dynamical scalars which respects $\frac{E_{7(7)}}{SU(8)}$ quotient symmetry and show that this is equivalent to (\ref{L2})\footnote{\ Although $E_{7(7)}$ is a non-compact (semisimple)group, it has been shown in \cite{Cahill:1978ps} that this construction leads a Lagrangian \textit{without} ghosts.}. First, we start with a general group element $\cV$ of $E_{7(7)}$ which depends on all 133 scalar fields. To reduce the number of scalars down to dimension of the coset space $\frac{E_{7(7)}}{SU(8)}$, we define an equivalence relation in the following way
\begin{eqnarray}\label{equiv}
\cV\sim\cV'\ ,\quad \mbox{if}\ \ \exists\ \ {\mathcal{U}}(y,\bar{y})\in SU(8)\ ,\quad\mbox{such that}\ \cV'={\mathcal{U}}\cV\ .
\end{eqnarray}
In other words, any two elements of $E_{7(7)}$ are equivalent if they differ by an $SU(8)$ multiplication from left. Now, by imposing this equivalence relation, the number of scalars reduces to $133-63=70$. In order to have (\ref{Vtrans}) symmetry (which is equivalent to (\ref{deltay}) and (\ref{Uinfi}) in the infinitesimal form), we also require that the Lagrangian must be invariant under the following global transformation
\begin{eqnarray}\label{Vright}
\cV\longrightarrow \cV'=\cV E^{-1}\ ,
\end{eqnarray}
where $E\in E_{7(7)}$. Now, in order to construct the Lagrangian, we introduce a gauge field connection $\cQ_{\mu}$ which belongs to $\mathfrak{su}(8)$ (Lie algebra of the local $SU(8)$ group) and its transformation law under local $SU(8)$ group is defined as
\begin{eqnarray}\label{Qtrans}
\cQ_{\mu}\longrightarrow {\mathcal{U}}\cQ_{\mu}{\mathcal{U}}^{-1}-(\partial_{\mu}{\mathcal{U}}){\mathcal{U}}^{-1}\ .
\end{eqnarray}
One is then able to construct a covariant derivative via the above connection as
\begin{eqnarray}\label{cov}
D_{\mu}\cV\equiv \partial_{\mu}\cV-\cQ_{\mu}\cV\ .
\end{eqnarray}
It is now evident that the covariant derivative (\ref{cov}) transforms homogeneously under local $SU(8)$ transformation
\begin{eqnarray}\label{covtrans}
D_{\mu}\cV\longrightarrow {\mathcal{U}}(D_{\mu}\cV)\ .
\end{eqnarray}
Using the above property, we can construct the Lagrangian (\ref{lag2}) which is invariant both under (\ref{Vright}) and the $SU(8)$ local transformations. Note that the above Lagrangian is still in terms of all 133 scalar fields of $E_{7(7)}$. However, we realize that the existence of $\cQ_{\mu}$ is pure algebraic and there is no derivative of $\cQ_{\mu}$ in (\ref{lag2}). Therefore, it can be eliminated using its equation of motion. Varying $\cQ_{\mu}$ and keeping $\cV$ fixed, we obtain the following relation
\begin{eqnarray}\label{equQ}
\delta{\mathcal{L}}=\mbox{Tr}\Bigg[\delta\cQ^{\mu}\big((\partial_{\mu}\cV)\cV^{-1}-\cQ_{\mu}\big)\Bigg]=0\ .
\end{eqnarray}
Since both $\cQ_{\mu}$ and $\delta\cQ_{\mu}$ belong to $\mathfrak{su}(8)$, (\ref{equQ}) implies that
\begin{eqnarray}\label{Pmu}
\cP_{\mu}\equiv(\partial_{\mu}\cV)\cV^{-1}-\cQ_{\mu}=(D_{\mu}\cV)\cV^{-1}\ ,
\end{eqnarray}
belongs to the \textit{orthogonal complement} of $\mathfrak{su}(8)$ in the Lie algebra of $E_{7(7)}$, namely the Lie algebra associated with the quotient space. This shows explicitly why $\cP_{\mu}$ belongs to the orthogonal complement of $\mathfrak{su}(8)$. Therefore, if we substitute the connection $\cQ_{\mu}$ by its algebraic equation of motion, we find a Lagrangian in terms of only 70 scalar fields which respects both local $SU(8)$ symmetry and the global symmetry (\ref{Vright}). Using (\ref{Pmu}), the Lagrangian in terms of 70 $y_{ijkl}$ scalars has the following form
\begin{eqnarray}\label{lag3}
{\mathcal{L}}=-\frac{1}{2}\mbox{Tr}\big(\cP^{\mu}\cP_{\mu}\big)\ .
\end{eqnarray}
The $\cQ_{\mu}$ appearance of (\ref{lag2}) is, of course, artificial and it is eliminated by its equation of motion. This proves that that the first term in (\ref{L2}) is invariant under the desired transformation.

\subsection{The ${\mathcal{L}}_{3}$ Part}

The $ {\cal L}_3$ part of the action as given in (\ref{L3}) requires some reorganization \cite{de Wit:1982ig}, to prove the invariance under (\ref{deltaF+})  together with the corresponding transformations of other fields. One rewrites the terms depending on vectors and the quartic fermion terms as follows
\be
{\cal L}_3 = -{1\over 8} e F_{\mu\nu IJ}^+ \, G^{+\mu\nu}{}_{IJ} - {1\over 4} \O^{+ij}_{\mu\nu}\bar F^{+\mu\nu} {}_{ij} +\mbox{h.c.}\ ,
\label{vectors}\ee
where the new symbol $\bar F^{+\mu\nu} {}_{ij}$ is $SU(8)$ covariant. It is defined via
\begin{eqnarray}\label{barF}
\cV \, \left(
  \begin{array}{c}
    F^{+}_{1\mu\nu} \\
    F^{+}_{2\mu\nu} \\
  \end{array}
\right)= \left(
  \begin{array}{c}
    \bar F^{+}_{\mu\nu ij} \\
   \O^{+}_{\mu\nu}{}^{ij} \\
  \end{array}
\right)\ .
\end{eqnarray}
The first term in (\ref{vectors}) vanishes by partial integration with account of exact equations of motion
\be
\partial_\mu[ e(G^{+\mu\nu}{}_{IJ} + G^{-\mu\nu  IJ})]=0\ .
\ee
The second term in (\ref{vectors}) is a product of $SU(8)$ tensors. This accomplishes the final step in the proof that the total Largangian after gauge-fixing is invariant under the non-linearly realized $E_{7(7)}$ when equations of motion of the theory are satisfied.


\end{document}